\documentclass[11pt,a4paper]{article}
\usepackage[utf8]{inputenc}
\usepackage{graphicx}
\usepackage{subcaption}
\usepackage{amsmath,amssymb,mathtools}
\usepackage{hyperref}
\hypersetup{colorlinks=true,linkcolor=blue,citecolor=blue,urlcolor=blue}
\usepackage{geometry}
\date{}
\begin{document}
\title{{\bf Affine quantization of the dynamical Reissner--Nordström region}}

\author{Morteza Bajand\thanks{%
e-mail: m.bajand@iau.ac.ir}\,\, and Babak Vakili\thanks{email:
ba.vakili@iau.ac.ir (corresponding author)}\\\\{\small {\it
Department of Physics, CT.C., Islamic Azad
University, Tehran, Iran}}} \maketitle

\begin{abstract}
We study the quantum dynamics of the dynamical region of the
Reissner--Nordström geometry using a minisuperspace reduction and affine
quantization, which is naturally suited for positive-definite geometrical
variables.
The resulting Wheeler--DeWitt equation becomes separable, yielding
Hermite-polynomial modes in one sector and Gaussian-like radial solutions in
the other.
Affine quantization introduces additional short-distance contributions that
modify the small-radius behaviour of the wave function.
By constructing normalizable semiclassical wave packets, we analyze the
resulting probability distributions in minisuperspace and the role played by
the electric charge in the quantum dynamics.
Our results extend previous affine-quantization studies of the Schwarzschild
case to the charged Reissner--Nordström geometry.
\vspace{5mm}\noindent\\
\newline Keywords: Affine quantization; Reissner--Nordström black hole; Singularity avoidance; Quantum black hole 
\end{abstract}

\section{Introduction}

Black holes remain among the most profound predictions of general relativity, providing a concrete arena where classical gravitational dynamics is
pushed to its limits. The simplest static and spherically symmetric vacuum
solution is the Schwarzschild geometry, whose interior region inevitably leads
to a classical curvature singularity. When electric charge is included, the
resulting Reissner--Nordström (RN) metric exhibits a richer causal structure:
an outer event horizon $r=r_{+}$, an inner Cauchy horizon $r=r_{-}$, and an
ingoing timelike singularity located at $r=0$. The presence of two distinct
horizons partitions the spacetime into multiple regions with alternating
$(T,R)$ roles of the coordinates, and the interior region ($r_{-}<r<r_{+}$)
acquires a cosmological character. Although the RN solution is classically
well understood, its inner horizon and timelike singularity continue to present
important conceptual challenges related to predictability, stability and the
ultimate fate of infalling observers
\cite{misner1973gravitation,poisson1990internal,brady1995critical}.

From a quantum-gravity perspective, the strong-field regime inside black holes
is expected to receive important quantum corrections beyond the classical
description. Minisuperspace quantization of black-hole interiors, inspired by
the Wheeler--DeWitt (WDW) approach to canonical quantum gravity, has therefore
attracted considerable attention
\cite{kuchar1994geometrodynamics,bojowald2005nonsingular,perez2015black}.
In many approaches, the dynamical region of the geometry is modeled as an
effective homogeneous system whose physical degrees of freedom reduce to a
finite number of time-dependent variables
\cite{QBH1}-\cite{QBH7}. Quantization of such reduced models, however,
often encounters technical difficulties related to operator self-adjointness
and the treatment of positive-definite configuration variables, such as the
areal radius or scale-factor--like degrees of freedom.
Establishing mathematically consistent operator domains and obtaining
well-defined quantum dynamics therefore become nontrivial tasks.

Affine quantization, originally formulated by Klauder and further developed in
various contexts \cite{Aff9}-\cite{klauder2012enhanced}, provides a
mathematically natural framework for systems whose configuration variables are
restricted to positive values. Instead of quantizing the canonical pair
\((q,p)\) for \(q>0\), one promotes the affine variables
\((q,d=qp)\) to quantum operators satisfying the affine commutation relation
\([\hat q,\hat d]=i\hbar\,\hat q\).
This algebra admits a self-adjoint representation on
\(L^{2}(\mathbb{R}_{+},dq/q)\) and automatically preserves the positivity of
the geometrical variable \(q\).
When implemented through affine coherent states, the resulting quantum
Hamiltonian acquires additional \(\hbar^{2}\)-dependent contributions,
including a characteristic term proportional to \(1/q^{2}\), which modifies
the short-distance dynamics of the system.
Recent applications to black-hole minisuperspace models, including the
Schwarzschild case, have shown that affine quantization provides a consistent
framework for obtaining well-defined quantum states and regular wave-function
behaviour in the small-\(q\) regime
\cite{AffineSchw}.

In this work we extend the affine quantization program to the
dynamical region of the RN geometry lying between the two
horizons. Starting from a minisuperspace reduction of this homogeneous sector,
we construct the corresponding affine Hamiltonian and derive the associated
WDW equation governing the quantum dynamics.
The resulting WDW equation admits an exact separation of variables:
one sector reduces to a harmonic-oscillator problem with Hermite-polynomial
eigenfunctions, while the radial sector is governed by an effective potential
containing both the affine contribution and the charge-dependent term.
Using these mode functions, we construct normalizable semiclassical wave
packets and analyze their behaviour in minisuperspace.
Our results show that the affine framework significantly modifies the
small-radius regime of the quantum dynamics and provides a mathematically
consistent description of positive-definite geometrical variables in the
charged black-hole case.

The paper is organized as follows.
In sections 2 and 3, we review the classical
RN geometry and construct the corresponding minisuperspace
model for the dynamical region between the two horizons.
Section 4 summarizes the affine quantization framework and applies it to the
reduced RN Hamiltonian.
In section 5, we derive and analyze the associated WDW equation,
while section 6 is devoted to the construction of semiclassical wave packets
and their behaviour in minisuperspace.
Finally, section 7 contains concluding remarks and discusses the physical
implications of the results.

\section{Reissner--Nordström geometry}
\label{sec:RN_classical}

Before constructing the minisuperspace model and performing affine quantization,
it is useful to recall the main geometric features of the RN
black hole and to clarify which region of the spacetime is quantized and why.
The RN solution represents the unique static and spherically symmetric solution of the
Einstein--Maxwell equations for a charged source with mass $M$ and electric charge $Q$.
Its line element in Schwarzschild-like coordinates reads

\begin{equation}
\label{RN_metric}
ds^2 = - f(r)\, dt^2 + \frac{dr^2}{f(r)} + r^2 d\Omega^2,
\qquad
f(r)=1-\frac{2M}{r}+\frac{Q^2}{r^2},
\end{equation}
where $d\Omega^2$ denotes the metric on the unit 2-sphere. For sub-extremal black holes ($|Q|<M$), the function $f(r)$ has two distinct positive
roots

\begin{equation}
r_\pm = M \pm \sqrt{M^2 - Q^2},
\end{equation}
corresponding respectively to the outer (event) horizon $r_+$ and the
inner (Cauchy) horizon $r_-$. These horizons split the spacetime into three
qualitatively different regions:
(i) Region I ($r>r_+$): the exterior, static and asymptotically flat region,
where $t$ is timelike and $r$ is spacelike,
(ii) Region II ($r_-<r<r_+$): a non-static region between the two horizons,
where the roles of $t$ and $r$ are interchanged, so that the coordinate $r$
becomes timelike and infalling observers inevitably move toward decreasing $r$ and
(iii) Region III ($0<r<r_-$): the innermost region beyond the Cauchy horizon,
where $f(r)$ becomes positive again and the coordinate $r$ reverts to being spacelike.
The timelike singularity at $r=0$ lies in this region.

In the present work we restrict our attention to Region II, namely the dynamical
region between the two horizons where the radial coordinate behaves as a timelike
variable and a homogeneous minisuperspace description becomes possible. In this
region one has $f(r)<0$, and the roles of $t$ and $r$ interchange. Introducing
the interior time coordinate $\tau$ via

\begin{equation}
\tau = r,
\qquad \text{with} \qquad
f(\tau)<0,
\end{equation}
and defining the positive function

\begin{equation}
F(\tau)\equiv -f(\tau)
=
\frac{2M}{\tau}-1-\frac{Q^2}{\tau^2}>0,
\end{equation}
the RN metric takes the explicitly time-dependent form

\begin{equation}
\label{RN_interior_metric}
ds^2
=
-\frac{d\tau^2}{F(\tau)}
+
F(\tau)\,d\chi^2
+
\tau^2 d\Omega^2,
\end{equation}
where $\chi$ is a spacelike coordinate replacing $t$ in the interior.
The metric \eqref{RN_interior_metric} has $\mathbb{R}\times S^2$ spatial topology
and is of anisotropic Kantowski--Sachs type. The dynamical variable $\tau$
therefore plays the role of an effective scale factor for the homogeneous
interior geometry between the two horizons.

This dynamical interior region provides the starting point for our minisuperspace
construction and for the affine quantization of the positive geometrical variable
$h=\tau^2$. The maximal analytic extension of RN contains additional asymptotic
regions together with an infinite chain of black-hole and white-hole domains.
However, the present analysis is restricted to the homogeneous region between
the horizons, where the minisuperspace reduction remains valid.

The spacetime possesses a timelike curvature singularity at $r=0$, where the curvature
invariants diverge, for example

\begin{equation}
R_{\mu\nu\rho\sigma}R^{\mu\nu\rho\sigma}
=
\frac{48 M^2}{r^6}
-
\frac{96 MQ^2}{r^7}
+
\frac{56 Q^4}{r^8}
\;\longrightarrow\;
\infty
\qquad
\text{as }
r\to 0.
\end{equation}
Unlike the Schwarzschild case, the RN spacetime also contains an inner Cauchy
horizon at $r_-$, which is known to suffer from classical instabilities such as
mass inflation. These features strongly suggest that the classical description
of the geometry becomes unreliable near the inner horizon and in the deep interior
region. The purpose of the present work is therefore not to quantize the full
maximally extended RN spacetime, but rather to investigate how affine
quantization modifies the effective dynamics of the homogeneous region between
the two horizons.

By focusing on this dynamical sector, the radial degree of freedom can be treated
as the effective scale factor of a minisuperspace model. This is precisely the
setting in which affine quantization becomes particularly useful: the relevant
geometrical variable (the areal radius $r$ or its square $h=r^2$) is strictly
positive, and affine quantization naturally preserves this positivity while
producing a well-defined kinetic operator supplemented by a repulsive
short-distance contribution proportional to $1/h^2$.

\section{Classical phase space for the RN dynamical interior}
We start from the Einstein--Maxwell action (in units $c=1$)

\begin{equation}\label{eq:EMaction}
S \;=\; \frac{1}{16\pi G}\int d^4x\sqrt{-g}\,{\cal R} \;-\; \frac{1}{4}\int d^4x\sqrt{-g}\,F_{\mu\nu}F^{\mu\nu} \;+\; S_{\rm bnd}.
\end{equation}
For the dynamical region between the two horizons (Region II), where the radial
coordinate behaves as a timelike variable, we adopt the following homogeneous,
spherically symmetric ansatz
\begin{equation}\label{eq:ansatz}
ds^2 \;=\; -N^2(\tau)\,d\tau^2 \;+\; a(\tau)\,dr^2 \;+\; h^2(\tau)\,d\Omega^2,
\qquad h(\tau)>0,\; a(\tau)>0.
\end{equation}
The coordinate volume of the spatial slices is denoted by $V_0$ (coming from the $r$-interval
and $\int d\Omega=4\pi$), identical in role to the Schwarzschild analysis of Ref.~\cite{AffineSchw}.

For a purely radial electric field the only independent component of the Maxwell tensor is
$F_{\tau r}$, and Gauss' law gives a conserved electric flux proportional to the charge $Q$.
Explicitly,
\begin{equation}
Q = \int_{S^2}\sqrt{-g}\,F^{r\tau}d\Omega
   = -\frac{4\pi h^2}{N\sqrt{a}}\,F_{\tau r},
\end{equation}
hence
\begin{equation}
F_{\tau r} = -\frac{N\sqrt{a}}{4\pi h^2}\,Q .
\end{equation}
The Maxwell action, after integrating over $S^2$ and the $r$-interval, reduces to\footnote{At this stage we perform an admissible redefinition of the lapse function,
which is possible due to the gauge nature of $N$. This absorbs the
nonessential factors of $a$ appearing in the Maxwell sector and leads
to a simplified minisuperspace Hamiltonian constraint.}

\begin{equation}
L_{\rm EM,red}
=\,N(\tau)\,V_0\,\frac{\mathcal{C}_{\rm EM}Q^2}{h^2(\tau)},
\label{LEMfinal}
\end{equation}
where $V_0$ denotes the (constant) angular volume factor (in practice $V_0=4\pi$ or a regulated finite factor) and $\mathcal{C}_{\rm EM}=\frac{1}{32\pi^2}$ is a calculable numerical factor arising from the metric prefactors. This fixes the model-independent minisuperspace scaling of the electric potential as $Q^2/h^2$.

The nonvanishing Christoffel symbols for \eqref{eq:ansatz} lead to the Ricci scalar

\begin{equation}\label{Ricci}
{\cal R}=\frac{2}{N^2}\!\left[
\frac{\ddot a}{2a} + 2\frac{\ddot h}{h}
-\frac{\dot N}{N}\!\left(\frac{\dot a}{2a}+2\frac{\dot h}{h}\right)
-\frac{\dot a^2}{4a^2} -\frac{\dot h^2}{h^2} -\frac{\dot a\,\dot h}{a h}
\right]
+ \frac{2}{h^2}.
\end{equation}
After integrating over angular coordinates and collecting terms, the gravitational
minisuperspace Lagrangian takes the schematic form

\begin{equation}\label{Lgrav}
L_{\rm grav,red}
= \frac{V_0}{16\pi G}
\left[
\frac{1}{N}\big(\alpha_{11}\dot h^2 + 2\alpha_{12}\dot h\dot a + \alpha_{22}\dot a^2\big)
- N\,\mathcal{V}_{\rm grav}(h,a)
\right],
\end{equation}
where $\alpha_{ij}$ and $\mathcal{V}_{\rm grav}$ are the explicit polynomials arising from \eqref{Ricci}.
Their detailed expressions can be generated algebraically but are not essential at this stage. So, the total minisuperspace Lagrangian can be written as

\begin{equation}\label{eq:Lred}
L_{\rm red} \;=\; \frac{V_0}{16\pi G}\left[\frac{1}{N}\big( \alpha_{11}\dot h^2 + 2\alpha_{12}\dot h\dot a + \alpha_{22}\dot a^2\big) - N\,\mathcal{V}_{\rm grav}(h,a)\right] - N V_0\frac{\mathcal{C}_{\rm EM}Q^2}{h^2}.
\end{equation}
To diagonalize the kinetic part we use the linear combinations

\begin{equation}
x=\frac{h+\mu a}{2},
\qquad
y=\frac{h-\mu a}{2},
\end{equation}
with $\mu$ chosen to remove cross-terms in the kinetic form.  
As in the Schwarzschild case, we introduce

\begin{equation}
u=\sqrt{x+y}=\sqrt{h}>0,
\qquad
v=\frac{x-y}{\sqrt{x+y}} = \frac{\mu a}{u}\in\mathbb{R}.
\end{equation}
This leads to a diagonal kinetic term $\propto (\dot u^2 - \dot v^2)$. The inverse relations read 

\begin{equation}
h=u^2,
\qquad
a=\frac{u\,v}{\mu}.
\end{equation}
Now, performing the Legendre transform and introducing the conjugate momenta $p_u,p_v$, the Hamiltonian constraint takes the compact form

\begin{equation}\label{eq:Hfinal}
\mathcal{H} \;=\; -\tilde N\Bigg[\frac{1}{4M_{\rm Pl}^2V_0}\big(p_u^2-p_v^2\big) - V_0M_{\rm Pl}^2(u^2-v^2) + \Gamma\,\frac{Q^2}{(u-v)^2}\Bigg] \;=\;0, 
\end{equation}
where $M_{\rm Pl}^2=(8\pi G)^{-1}$ and the coefficient $\Gamma$ is given explicitly by

\begin{equation}
\Gamma
= \mathcal{C}_{\rm EM}\,\frac{1}{(\partial h/\partial(u-v))^2}
= \frac{1}{32\pi^2}
\,\frac{1}{(\partial h/\partial(u-v))^2},
\end{equation}
i.e.\ it is wholly determined once the explicit $u,v\!\leftrightarrow\! h,a$ map is fixed. Since \(u>0\), the affine quantization naturally preserves the positivity
of the geometrical variable throughout the dynamical region. The resulting
affine representation also improves the behaviour of the quantum operators
in the small-\(u\) regime. Using the parametrization $h=u^2$, the Hamiltonian constraint finally takes the form

\begin{equation}\label{Cl Ham}
\mathcal{H} = -\tilde N\!\left[\frac{1}{4M_{\rm Pl}^2 V_0}(p_u^2-p_v^2) - V_0 M_{\rm Pl}^2 (u^2 - v^2)
+ \mathcal{C}_{\rm EM}\,\frac{Q^2}{u^4}\right]=0,
\end{equation}
and provides the starting point for the affine quantization of the
homogeneous dynamical sector of the RN geometry. The variable
$u$ ranges over $(0,\infty)$ while $v\in\mathbb R$.

\section{A brief review of affine quantization}

Standard canonical quantization is naturally adapted to phase–space variables
\((q,p)\in\mathbb{R}^2\), with no restriction on the configuration coordinate.
However, when the classical phase space contains a positively defined
variable, $q>0$, the canonical representation fails to provide a self–adjoint
momentum operator on the physical domain \cite{Aff1}-\cite{Aff3}. This difficulty appears in many
gravitational minisuperspace models, including black hole interiors, where the
radial or scale–factor–like variables never cross zero. Affine quantization
provides a consistent alternative strategy, ensuring the correct operator
domains and yielding well–defined semi–classical corrections \cite{Aff4}-\cite{Aff8}.

Let $q>0$ be the positive configuration variable of the classical phase space.
Instead of the canonical pair $(q,p)$, affine quantization employs the
affine pair

\begin{equation}
(q,d), \qquad d \equiv pq.
\end{equation}
The non–linear transformation $p\mapsto d=pq$ encodes the positivity of $q$
directly into the algebra.  The fundamental Poisson bracket reads

\begin{equation}
\{q,d\}=q,
\end{equation}
which leads upon quantization to the affine commutation relation

\begin{equation}
[\hat q,\hat d] = i\hbar \hat q.
\label{affineCCR}
\end{equation}
This algebra admits an essentially unique irreducible self–adjoint
representation on the Hilbert space

\begin{equation}
\mathcal{H}=L^2(\mathbb{R}_+, dq/q),
\end{equation}
where the operators act as

\begin{equation}
(\hat q\psi)(q)=q\,\psi(q),\qquad
(\hat d\psi)(q)=-i\hbar\left(q\frac{d}{dq}+\frac12\right)\psi(q).
\end{equation}
Both operators are self–adjoint on suitable dense domains, and in particular
$\hat q$ has spectrum $\mathbb{R}_+$, as required.

Following the construction initiated in the affine program in Refs.~ \cite{Aff9, Aff10, klauder2012enhanced, Aff11}, one introduces affine
coherent states

\begin{equation}
|p,q\rangle
= e^{\frac{i}{\hbar}p\hat q}\,
e^{-\frac{i}{\hbar}\ln(q)\hat d}\,
|\eta\rangle,
\end{equation}
where $|\eta\rangle$ is a fiducial vector chosen to satisfy an admissibility
condition ensuring a resolution of the identity. These states provide a natural
semi–classical phase–space representation adapted to $q>0$:

\begin{equation}
\int_{\mathbb{R}}\!\int_0^\infty
\frac{dp\,dq}{2\pi\hbar\,q}\,
|p,q\rangle\langle p,q|=\mathbb{I}.
\end{equation}
Given any classical function $f(q,p)$, one associates its \emph{lower symbol}
(semi–classical counterpart)

\begin{equation}
f^\sharp(q,p)
=\langle p,q|\,\hat f\,|p,q\rangle,
\end{equation}
where $\hat f$ is the operator obtained by symmetric ordering with respect to
the affine pair $(\hat q,\hat d)$.  The resulting $f^\sharp$ differs from the
classical $f$ by $\hbar^2$ corrections that encode the affine geometry.  A
characteristic example is the kinetic term

\begin{equation}
p^2 \quad\longrightarrow\quad
p^2 + \frac{\gamma}{q^2},
\end{equation}
where the coefficient $\gamma>0$ depends on the choice of fiducial vector.
This correction acts as an effective repulsive potential that strongly modifies
the short-distance dynamics near $q=0$ and improves the behaviour of the
corresponding quantum operators.

In many gravitational reduction schemes the interior geometry of a black hole
is described by a finite set of homogeneous degrees of freedom, at least one
of which (denoted here by $u$) is strictly positive.  Canonical quantization
would require $p_u=-i\hbar\partial_u$ on $L^2(\mathbb{R}_+,du)$, which is not
self–adjoint and suffers from domain–extension ambiguities.  Affine
quantization avoids these difficulties by promoting the pair $(u,p_u)$ to the
affine variables $(u,d_u=u p_u)$ obeying the commutation rule
$[\hat u,\hat d_u]= i\hbar\hat u$ and represented on $L^2(\mathbb{R}_+,du/u)$.
The corresponding affine coherent–state quantization yields a semi–classical
Hamiltonian

\begin{equation}
H^\sharp(u,p_u) = H_{\rm cl}(u,p_u)\,+\,\Delta H_{\rm aff}(u),
\end{equation}
in which the correction term
$\Delta H_{\rm aff}(u)\propto \hbar^2/u^2$
generically appears. This additional positive contribution significantly
modifies the quantum dynamics in the small-$u$ regime and tends to suppress
the concentration of wave packets near the classical boundary $u=0$.

The dynamical region of the RN geometry between the two horizons,
once reduced to a minisuperspace form,
contains a positive scale–factor–like variable $u(t)>0$
governing the radial two–sphere. Affine quantization thus provides a natural and mathematically
consistent framework for quantizing the interior geometry. The resulting WDW equation inherits a repulsive $\sim 1/u^2$
contribution, modifying the effective dynamics in the small-$u$ regime
as well as the associated quantum wave propagation. These properties have already been demonstrated in related
black–hole minisuperspace analyses, including the Schwarzschild case, and play a central role in the present extension to the dynamical
sector of the RN geometry.

\section{Affine quantization of the RN dynamical sector}
\label{sec:affine_quant}
In the reduced phase space describing the homogeneous dynamical region
between the two RN horizons, the variable
\(u\in(0,\infty)\) is strictly positive, while \(v\in\mathbb{R}\) is unrestricted.
Affine quantization is therefore naturally applied to the \((u,p_u)\)-sector, 
while the \((v,p_v)\)-sector is quantized canonically. According to the affine quantization rules, one replaces the canonical pair \((u,p_u)\) by the operators

\begin{equation}
\hat{u}>0,\qquad 
\hat{d}=\frac{1}{2}\left(\hat{u}\,\hat{p}_u+\hat{p}_u\,\hat{u}\right),
\end{equation}
which obey the affine commutator

\begin{equation}
[\hat{u},\hat{d}] = i\hbar\,\hat{u}.
\end{equation}
The canonical momentum operator does not exist as a self-adjoint operator on
the Hilbert space \(L^2(\mathbb{R}^+,du)\), while the dilation operator \(\hat{d}\) does.
A convenient representation is

\begin{equation}
\hat{u}\,\psi(u)=u\,\psi(u),\qquad
\hat{d}\,\psi(u)=-i\hbar\left( u\frac{d}{du}+\frac{1}{2}\right)\psi(u).
\end{equation}
The quadratic kinetic term \(p_u^2\) has the affine image

\begin{equation}
\label{affine_kinetic}
\widehat{p_u^2}
\;\longrightarrow\;
-\hbar^2\left(
\frac{d^2}{du^2} - \frac{\kappa}{u^2}
\right),
\end{equation}
where the constant \(\kappa>0\) encodes the affine ``centrifugal'' term.
Using the fiducial vector employed in the Schwarzschild case \cite{AffineSchw}, one obtains the explicit value $\kappa=\frac{3}{4}$. Thus,

\begin{equation}
\label{A_u_operator}
\widehat{p_u^2}\,\psi(u)
=-\hbar^2\left[
\psi''(u) - \frac{3}{4u^2}\psi(u)
\right].
\end{equation}
The \(v\)-sector is quantized canonically in the usual representation

\begin{equation}
\hat{v}\psi(v)=v\psi(v),\qquad \hat{p}_v=-i\hbar\,\partial_v,\qquad \widehat{p_v^2}=-\hbar^2\frac{\partial^2}{\partial v^2}.
\end{equation}
Now, recalling the classical Hamiltonian constraint (\ref{Cl Ham}), we promote each sector to operators using the affine and canonical rules.
The quantum Hamiltonian constraint \(\widehat{\mathcal{H}}\Psi=0\) becomes

\begin{equation}\label{WDW_final}
\Bigg[\frac{\partial^2}{\partial v^2}-\left(\frac{\partial^2}{\partial u^2}-\frac{3}{4u^2}\right)\;+\;\Omega^4\,(u^2 - v^2)\;-\;\Xi\,\frac{Q^2}{u^4}\Bigg]\Psi(u,v)=0,
\end{equation}
where we abbreviated

\[
\Omega^4 \equiv 
\frac{4M_{\rm Pl}^4 V_0^2}{\hbar^2},
\qquad
\Xi\equiv
\frac{4M_{\rm Pl}^2 V_0^2}{\hbar^2}\,\mathcal{C}_{\rm EM}.
\]
The potential term

\begin{equation}
V_{\rm eff}(u,v)=\Omega^4(u^2-v^2)-\Xi\frac{Q^2}{u^4}+\frac{3}{4u^2},
\end{equation}
shows explicitly that the affine contribution \(+3/(4u^2)\)
significantly modifies the effective dynamics in the small-\(u\) regime,
while the electromagnetic term introduces an additional strong
short-distance contribution proportional to \(Q^2/u^4\). If we introduce a natural length scale

\begin{equation}
u_0 \;=\; \frac{1}{\Omega},
\end{equation}
and define dimensionless coordinates

\begin{equation}
x\equiv \frac{u}{u_0},\qquad y\equiv \frac{v}{u_0},
\end{equation}
the dimensionless WDW equation reads

\begin{equation}\label{WDW_dimless}
\Bigg[
\frac{\partial^2}{\partial y^2}
-\Big(\frac{\partial^2}{\partial x^2}-\frac{3}{4x^2}\Big)
+ (x^2-y^2)
- \lambda\frac{1}{x^4}
\Bigg]\Psi(x,y)=0,
\end{equation}
where we defined the single dimensionless coupling
\begin{equation}\label{lambda_def}
\lambda \;\equiv\; \Xi\,\Omega^2\,Q^2\;,
\end{equation}
i.e.\ all physical constants are collected into \(\lambda\) and if desired one can substitute \(\Xi\) and \(\Omega\) from their definitions
\(\Omega^4=4M_{\rm Pl}^4 V_0^2/\hbar^2\) and
\(\Xi=4M_{\rm Pl}^2 V_0^2\mathcal{C}_{\rm EM}/\hbar^2\) to obtain an explicit formula
for \(\lambda\) in terms of \(M_{\rm Pl},V_0,\hbar,\mathcal{C}_{\rm EM}\) and \(Q\).

To Separate the variables, assume a product ansatz \(\Psi(x,y)=U(x)\,V(y)\). Substitute into \eqref{WDW_dimless} gives

\begin{equation}
\frac{V''(y)}{V(y)} - y^2 \;=\; \frac{U''(x)}{U(x)} -\frac{3}{4x^2} - x^2 + \frac{\lambda}{x^4}.
\end{equation}
Both sides equal a separation constant which we denote by \(-\varepsilon\). Thus we obtain

\begin{equation}\label{v_eq}
V''(y) + \big(\varepsilon - y^2\big)V(y)=0.
\end{equation}for the $V$-sector. This is the standard harmonic-oscillator equation. Square integrable
solutions exist for the discrete spectrum

\begin{equation}
\varepsilon_n = 2n+1,\qquad n=0,1,2,\dots,
\end{equation}
with normalized eigenfunctions given by Hermite--Gaussian functions

\begin{equation}\label{Vn}
V_n(y)=\frac{1}{\sqrt{2^n n!\sqrt{\pi}}}\,e^{-y^2/2}\,H_n(y).
\end{equation}
The corresponding equation for \(U(x)\) becomes in Schr\"odinger form

\begin{equation}\label{u_schrodinger}
\bigg[-\frac{d^2}{dx^2} + V_{\rm eff}(x) \bigg]U(x) \;=\; \varepsilon\,U(x),
\end{equation}
with the effective potential
\begin{equation}\label{Veff}
V_{\rm eff}(x) \;=\; \frac{3}{4x^2} + x^2 - \frac{\lambda}{x^4}.
\end{equation}
Thus the full spectrum is obtained by choosing pairs \((n,\;{\rm eigenvalue\;}\varepsilon)\)
such that \(\varepsilon=\varepsilon_n\), from the \(V\)-sector, and simultaneously \(\varepsilon\)
is an eigenvalue of the radial operator \eqref{u_schrodinger}. In practice we build standing-wave
solutions \(\Psi_{n,k}(x,y)=U_{k}^{(n)}(x)\,V_n(y)\) where \(U_{k}^{(n)}\) is the \(k\)-th
eigenfunction that satisfies \eqref{u_schrodinger} with eigenvalue \(\varepsilon=\varepsilon_n\). 

Before proceeding further, let us briefly discuss the qualitative behaviour
of the effective potential \(V_{\rm eff}(x)\).
The potential \eqref{Veff} contains three competing contributions:
(i) the affine term \(+3/(4x^2)\), which becomes important in the
small-\(x\) regime,
(ii) the confining oscillator term \(+x^2\), which dominates at large
\(x\) and supports normalizable solutions and
(iii) the electromagnetic contribution \(-\lambda/x^4\),
which strongly affects the short-distance structure of the WDW
equation.

The interplay between these terms determines the behaviour of the quantum
states near the boundary \(x\to0\). Depending on the value of \(\lambda\),
different spectral behaviours may arise. In practice, physically acceptable
(normalizable) solutions are obtained for a broad range of parameters,
and the affine contribution tends to suppress the concentration of the wave
function in the small-\(x\) region.

To obtain an approximate semi-analytic solution for the radial equation (\ref{u_schrodinger}), we employ a variational ansatz that is compatible with the affine structure in the small-\(x\) regime. A convenient family of trial functions is the \emph{power--Gaussian} class

\begin{equation}\label{trial_general}
U_{\alpha,s}(x) \;=\; \mathcal{N}_{\alpha,s}\; x^{s}\,e^{-\frac{\alpha x^2}{2}},
\qquad s>\tfrac{3}{2},\ \alpha>0,
\end{equation}
which guarantees finiteness of all expectation value appearing in the energy
functional for any \(s>\tfrac{3}{2}\). Using

\[
\int_0^\infty x^p e^{-\alpha x^2}\,dx=\tfrac12\alpha^{-(p+1)/2}\Gamma\!\Big(\tfrac{p+1}{2}\Big),
\]
the normalization constant is

\begin{equation}
\mathcal{N}_{\alpha,s}^2=\frac{1}{\displaystyle \frac12\alpha^{-(2s+1)/2}\Gamma\!\Big(\tfrac{2s+1}{2}\Big)}=\frac{2\,\alpha^{(2s+1)/2}}{\Gamma\!\Big(\tfrac{2s+1}{2}\Big)}.
\end{equation}
All matrix elements required for the variational energy
\(E[\alpha,s]=\langle U_{\alpha,s}|H|U_{\alpha,s}\rangle\) reduce to Gamma-ratios. Define for brevity

\begin{equation}
\Gamma_n \equiv \Gamma\!\Big(\tfrac{2s+n}{2}\Big),
\qquad n\in\{-3,-1,+1,+3\}.
\end{equation}
Then one obtains

\begin{eqnarray}\label{T_general}
\left\{
\begin{array}{ll}
T[\alpha,s] \equiv \langle U'|U'\rangle
=\alpha\; \frac{ s^2 \,\Gamma_{-1} + \Gamma_{+3} - 2s\,\Gamma_{+1} }{ \Gamma_{+1} },\\\\
\langle x^2\rangle= \alpha^{-1}\;\frac{\Gamma_{+3}}{\Gamma_{+1}},\\\\
\big\langle \tfrac{1}{x^2}\big\rangle= \alpha\;\frac{\Gamma_{-1}}{\Gamma_{+1}}, \\\\
\big\langle \tfrac{1}{x^4}\big\rangle= \alpha^{2}\;\frac{\Gamma_{-3}}{\Gamma_{+1}}.
\end{array}
\right.
\end{eqnarray}
Hence the variational energy is

\begin{equation}\label{E_general}
E[\alpha,s] \;=\; T[\alpha,s] \;+\; \frac{3}{4}\big\langle\tfrac{1}{x^2}\big\rangle \;+\; \langle x^2\rangle \;-\; \lambda \big\langle\tfrac{1}{x^4}\big\rangle,
\end{equation}
or explicitly

\begin{equation}\label{E Gamm}
E[\alpha,s]
= \frac{1}{\Gamma_{+1}}\Bigg\{ 
\alpha\Big( s^2\Gamma_{-1} + \Gamma_{+3} - 2s\Gamma_{+1}\Big)
\;+\;\frac{\Gamma_{+3}}{\alpha}
\;+\;\frac{3}{4}\,\alpha\,\Gamma_{-1}
\;-\;\lambda\,\alpha^2\,\Gamma_{-3}
\Bigg\}.
\end{equation}
The optimal variational parameter \(\alpha_{\rm opt}(k,\lambda)\) is obtained by
solving \(\partial_\alpha E[\alpha,s]=0\). In general, this leads to an algebraic equation in \(\alpha\), which can be solved analytically in
closed form or expanded perturbatively. A particularly convenient and convergent choice is \(s=2\). In this case the
relevant Gamma-values reduce to known constants

\[
\Gamma\!\Big(\tfrac{2s+1}{2}\Big)=\Gamma\!\Big(\tfrac{5}{2}\Big)=\tfrac{3\sqrt{\pi}}{4},\qquad
\Gamma\!\Big(\tfrac{2s+3}{2}\Big)=\Gamma\!\Big(\tfrac{7}{2}\Big)=\tfrac{15\sqrt{\pi}}{8},
\]
\[
\Gamma\!\Big(\tfrac{2s-1}{2}\Big)=\Gamma\!\Big(\tfrac{3}{2}\Big)=\tfrac{\sqrt{\pi}}{2},\qquad
\Gamma\!\Big(\tfrac{2s-3}{2}\Big)=\Gamma\!\Big(\tfrac{1}{2}\Big)=\sqrt{\pi}.
\]
Substituting these values into \eqref{E Gamm} yields the compact closed-form variational energy

\begin{equation}\label{E_k2}
E[\alpha] \;=\; \frac{5}{3}\,\alpha \;+\; \frac{5}{2}\,\frac{1}{\alpha} \;-\; \frac{4}{3}\,\lambda\,\alpha^2.
\end{equation}
The stationarity condition \(\partial_\alpha E=0\) gives

\begin{equation}\label{alpha_cubic}
16\,\lambda\,\alpha^3 - 10\,\alpha^2 + 15 \;=\; 0.
\end{equation}
This cubic can be solved exactly to give \(\alpha_{\rm opt}(\lambda)\), or handled by controlled approximations. For \(\lambda\ll1\), neglect the cubic term in \eqref{alpha_cubic} to obtain the leading value $\alpha_0 = \sqrt{\tfrac{3}{2}}$.
Including the first-order correction by linearizing around \(\alpha_0\) yields

\begin{equation}
\alpha_{\rm opt} \approx \alpha_0 + \delta,\qquad
\delta \approx \frac{16\,\lambda\,\alpha_0^3}{20\alpha_0} = \frac{4}{5}\,\lambda\,\alpha_0^2
= \frac{6}{5}\,\lambda.
\end{equation}
Hence for small \(\lambda\)

\begin{equation}
\alpha_{\rm opt} \approx \sqrt{\tfrac{3}{2}} \;+\; \tfrac{6}{5}\,\lambda \;+\; \mathcal{O}(\lambda^2). 
\end{equation}
Thus to first order in $\lambda$ one has a compact analytic approximation for $\alpha$,
easy to report and compare with numerics. If \(\lambda\) becomes large the \(-\lambda\alpha^2\) term tends to dominate the energy
and the optimum \(\alpha\) typically shifts. Balance of the leading powers in
\(\alpha\) suggests a scaling estimate; solving \(\partial_\alpha E\approx0\) with the
dominant contributions gives to leading order \(\alpha\sim \mathcal{O}(1/\lambda)\).
(One can obtain a more precise asymptotic by solving \eqref{alpha_cubic} in the
large-\(\lambda\) limit.) Substitute $\alpha_{\rm opt}$ into \eqref{E_k2} to obtain the variational
estimate $\varepsilon_{\rm var}=E[\alpha_{\rm opt}]$. Using the small-$\lambda$ expansion
one finds to first order

\begin{equation}
\varepsilon_{\rm var}(\lambda)\approx
E\!\Big(\sqrt{\tfrac{3}{2}}+\tfrac{6}{5}\lambda\Big)
= E_0 + E_1\,\lambda + \mathcal{O}(\lambda^2),
\end{equation}
where $E_0=E[\sqrt{3/2}]\approx
4.08248$ and $E_1=-2$ obtained by Taylor expansion. Thus, within the present variational approximation, the radial eigenvalue
decreases linearly with \(\lambda\) at first order.

\section{Wavefunction construction and physical implications}

The solutions of the affine WDW equation acquire physical meaning only after suitable wave packets are constructed and their probability distributions are analyzed. In this section we first build normalizable quantum states from the separated eigenmodes obtained in the previous section and investigate the corresponding probability densities in minisuperspace. Particular attention is paid to the influence of the charge parameter on the localization properties of the wave function and on the probability of approaching the classically singular region. We then discuss the physical interpretation of the resulting quantum states, their correspondence with the classical RN geometry and the implications of the affine quantum corrections for the event horizon, the Cauchy horizon and singularity avoidance.

\subsection{Wave packet construction and probability densities}

\begin{figure}[t]
\centering
\begin{subfigure}[b]{0.45\textwidth}
\centering
\includegraphics[width=\textwidth]{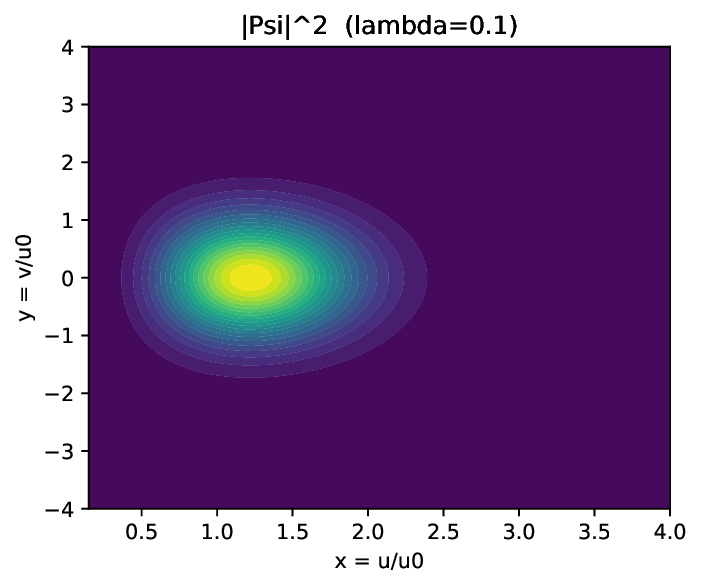}
\caption{$\lambda=0.10$}
\label{fig:psi010}
\end{subfigure}
\hfill
\begin{subfigure}[b]{0.45\textwidth}
\centering
\includegraphics[width=\textwidth]{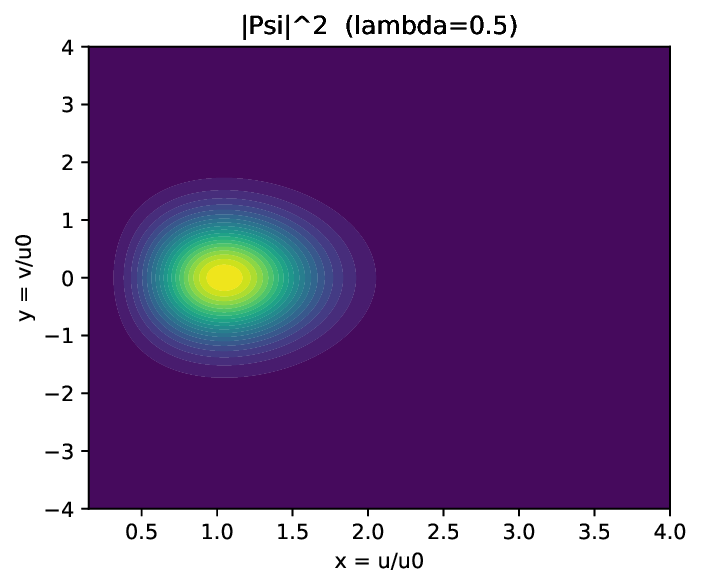}
\caption{$\lambda=0.50$}
\label{fig:psi005}
\end{subfigure}
\caption{Comparison of the normalized probability density
$|\Psi(x,y)|^2$ for two representative values of the
charge parameter $\lambda$.
As $\lambda$ increases, the peak of the probability distribution
moves toward smaller values of the minisuperspace radius $x$,
reflecting the stronger influence of the electromagnetic sector.
At the same time, the probability density remains strongly suppressed
near $x=0$, indicating that the affine quantum correction continues
to prevent localization at the classical singularity.}
\label{fig:psi_comparison}
\end{figure}

Starting from the separated product modes

\begin{equation}
\Psi_{n,k}(x,y)=U^{(n)}_k(x)\,V_n(y),
\end{equation}
where \(V_n(y)\) are given by Hermite-Gaussian functions (\ref{Vn}) with eigenvalues
\(\varepsilon_n=2n+1\) and \(U^{(n)}_k(x)\) are the radial eigenfunctions solving (\ref{u_schrodinger}) with the matching condition \(\varepsilon=\varepsilon_n\) for stationary product modes, we construct a physically motivated wavepacket as a superposition over the discrete \(V\)-modes \(n\) and, if desired, over radial labels \(k\):

\begin{equation}\label{wavepacket_general}
\Psi(x,y) \;=\; \sum_{n=0}^{\infty}\sum_{k=0}^{\infty} c_{n,k}\; U^{(n)}_k(x)\,V_n(y).
\end{equation}
To build a semiclassical packet peaked on a given \(V\)-quantum number \(n_0\) we take

\begin{equation}
c_{n,k} \;=\; c_n\,\delta_{k,0},\qquad
c_n \;=\; \mathcal{C}\,\exp\!\Big[-\frac{(n-n_0)^2}{2\sigma_n^2}\Big]\,e^{i\phi_n},
\end{equation}
where \(\sigma_n>0\) controls the spread in \(n\), \(\phi_n\) are optional phases (for e.g.
constructing a peaked semiclassical phase), and \(\mathcal{C}\) is the normalization constant
fixed by

\begin{equation}
1 \;=\; \int_0^\infty\!\int_{-\infty}^{\infty} |\Psi(x,y)|^2 \,dy\,dx
\;=\; \sum_{n=0}^\infty |c_n|^2 \underbrace{\int_0^\infty \big|U^{(n)}_{0}(x)\big|^2 dx}_{\equiv 1},
\end{equation}
where we have assumed the radial modes are individually normalized:
\(\int_0^\infty |U^{(n)}_0(x)|^2 dx = 1\) and \(\int_{-\infty}^{\infty}|V_n(y)|^2 dy = 1\).
Thus \(\mathcal{C}\) is simply

\begin{equation}
\mathcal{C} \;=\; \Bigg(\sum_{n=0}^\infty e^{-\frac{(n-n_0)^2}{\sigma_n^2}}\Bigg)^{-1/2}.
\end{equation}
To obtain the radial factors \(U^{(n)}_0(x)\), note that the exact radial eigenfunctions satisfying \(\varepsilon=\varepsilon_n=2n+1\) are obtained by
solving the Schr\"odinger problem (\ref{u_schrodinger}). In practice we may use the variational trial
\(U_{\alpha,s=2}(x)=\mathcal{N}\,x^{2}e^{-\alpha x^2/2}\) and determine \(\alpha=\alpha_n\)
by solving the matching condition

\begin{equation}\label{matching}
E[\alpha_n] \;=\; \varepsilon_n \;=\; 2n+1,
\end{equation}
where \(E[\alpha]\) is given in closed form by Eq. (\ref{E_k2}). For each \(n\) this yields a
positive \(\alpha_n\).
The normalized approximate radial mode is then\footnote{If for some \(n\) no positive solution \(\alpha_n\) exists in the variational ansatz,
one must compute \(U^{(n)}_0\) numerically by direct diagonalization of (\ref{u_schrodinger}).}

\begin{equation}
U^{(n)}_0(x) \approx U_{\alpha_n}(x) = \mathcal{N}(\alpha_n)\,x^{2}e^{-\frac{\alpha_n x^2}{2}},
\qquad
\mathcal{N}(\alpha) = \sqrt{\frac{2\,\alpha^{5/2}}{\Gamma(5/2)}}.
\end{equation}
To construct semiclassical wave packets localized around selected regions
of minisuperspace, one may choose phases \(\phi_n\) proportional to a classical action or simply \(\phi_n=0\).
If one wishes a packet localized in both \(x\) and \(y\) one can also sum a few radial
modes \(k\) with Gaussian weights in \(k\) as well. For example, take \(n_0\) (central \(V\)-mode), \(\sigma_n\) (width), \(\phi_n=0\), \(k=0\) only, and
compute for each \(n\) the \(\alpha_n\) from (\ref{matching}). Then, one may form the wave packet as 

\begin{equation}
\Psi(x,y)=\mathcal{C}\sum_{n=0}^{N_{\max}} e^{-(n-n_0)^2/(2\sigma_n^2)} U_{\alpha_n}(x)\,V_n(y),
\end{equation}
for which we may choose \(N_{\max}\approx n_0+4\sigma_n\) to capture most of the weight. Figure~\ref{fig:psi_comparison} shows the normalized radial
probability density obtained from the variational ground-state wave
function. The most important observation is that the probability density remains
strongly suppressed in the immediate vicinity of $x=0$ for all values
of $\lambda$ considered.

As the charge parameter increases, the maximum of the distribution
moves slightly toward smaller radii.
This behavior is consistent with the attractive character of the
electromagnetic contribution in the effective potential.
However, no accumulation of probability occurs at the classical
singularity.
Instead, the wave function continues to vanish as $x\rightarrow0$,
showing that the affine repulsive term remains sufficiently strong to
maintain regularity of the quantum state.

From a semiclassical viewpoint, the displacement of the probability
peak represents a charge-induced modification of the preferred quantum
configuration.
The effect is quantitative rather than qualitative:
the charge changes the localization properties of the state but does
not destroy the singularity-avoidance mechanism generated by affine
quantization.

With the normalized packet \(\Psi(x,y)\) one computes standard expectation values

\begin{equation}
\langle \mathcal{O} \rangle
=\int_0^\infty\int_{-\infty}^{\infty} \Psi^*(x,y)\,\widehat{\mathcal O}\,\Psi(x,y)\,dy\,dx.
\end{equation}
In particular the expectation value of the physical radius \(h = u^2 = (u_0 x)^2\) is

\begin{equation}
\langle h \rangle
= u_0^2 \sum_{n,k}\sum_{n',k'} c_{n,k}^* c_{n',k'} \int_0^\infty U^{(n)*}_{k}(x)\,x^2\, U^{(n')}_{k'}(x)\,dx
\times \int_{-\infty}^{\infty} V_n^*(y)V_{n'}(y)\,dy.
\end{equation}
Using orthonormality of the \(V_n\) modes the double sum collapses in \(n\). For the
recommended choice \(k=0\) this reduces to

\begin{equation}
\langle h \rangle = u_0^2 \sum_{n=0}^\infty |c_n|^2 \int_0^\infty x^2 |U^{(n)}_0(x)|^2 dx.
\end{equation}
For the trial with $s=2$ one finds

\begin{equation}
\langle x^2\rangle = \frac{\Gamma(7/2)}{\Gamma(5/2)}\frac{1}{\alpha} = \frac{5}{2}\frac{1}{\alpha},
\end{equation}
hence

\begin{equation}
\langle h \rangle = u_0^2\langle x^2\rangle = u_0^2\frac{5}{2\alpha}.
\end{equation}
Since $\alpha_{\rm opt}$ increases (weakly) with $\lambda$ for small $\lambda$, the mean radius $\langle h\rangle$ \emph{decreases} with increasing charge in this regime: the packet becomes more localized toward smaller $u$ (smaller physical radius).

To characterize the behaviour of the wave packet in the small-\(x\) regime,
we define the probability for the system to occupy the region \(x<\delta\)
(i.e.\ \(u<u_0\delta\)) as

\begin{equation}
\tilde P(\delta) \;=\;
\int_0^\delta\int_{-\infty}^\infty |\Psi(x,y)|^2\,dy\,dx
=
\sum_{n=0}^\infty |c_n|^2
\underbrace{\int_0^\delta |U^{(n)}_0(x)|^2 dx}_{\equiv p_n(\delta)}.
\end{equation}
With the variational trial \(U_{\alpha_n}\), the inner integrals can be
evaluated analytically

\begin{equation}
p_n(\delta)
=
\frac{\gamma\!\big(\tfrac{5}{2},\alpha_n \delta^2\big)}
{\Gamma\!\big(\tfrac{5}{2}\big)},
\end{equation}
where \(\gamma(s,z)\) denotes the lower incomplete Gamma function.
For small \(\delta\), the leading behaviour scales as
\(\sim (\alpha\delta^2)^{5/2}\), implying that larger values of
\(\alpha\) suppress the wave-function support in the small-\(x\) region.

Another direct consequence of the above analytic results concerns the role
of the charge parameter \(\lambda\).
Increasing \(\lambda\) enhances the contribution of the
electromagnetic term \(-\lambda/x^4\), thereby modifying the radial spectrum
and shifting the variational eigenvalues \(\varepsilon_{\rm var}\).
As a result, the spectral matching condition
\(\varepsilon_{\rm var}=\varepsilon_n\) is affected quantitatively by the
electric charge. At the same time, the affine contribution \(+3/(4x^2)\)
introduces an additional short-distance modification in the effective
potential.
The interplay between these two terms determines the behaviour of the
wave packet in the small-\(x\) regime. For sufficiently large values of \(\lambda\),
the variational approximation may cease to provide reliable results,
indicating the need for a full numerical treatment of the WDW
equation. Therefore, the present semi-analytic construction is expected to be most
accurate in the moderate-coupling regime.

It is important to emphasize that the present minisuperspace model does
not explicitly resolve the full global RN causal structure.
Consequently, the event horizon and the Cauchy horizon are not represented
as independent dynamical boundaries in the WDW equation.
Instead, their influence enters indirectly through the reduced
Hamiltonian governing the interior ($T$-region) geometry.

The probability-density plots therefore should not be interpreted as
local observables measured directly at the horizons.
Rather, they characterize the quantum state of the reduced interior
geometry and provide a quantitative measure of how the electric charge
modifies the localization properties of the wave function.

\subsection{Physical interpretation and classical correspondence}

The quantum solutions obtained in the present work admit a straightforward physical interpretation when compared with the classical RN geometry. In classical general relativity, the event horizon $(r=r_{+})$ is a regular null hypersurface rather than a curvature singularity. A freely falling observer crosses the horizon without encountering any locally divergent physical quantity. It is therefore important to verify that the quantum description preserves this basic feature of the classical spacetime.

The probability densities constructed from the WDW wave packets remain smooth throughout the dynamically accessible minisuperspace region. In particular, no divergence, discontinuity or anomalous suppression is observed in the region corresponding to the classical event horizon. The wave function evolves continuously across this region and the associated probability density does not display any pathological behavior. In this sense, the affine-quantized model is fully consistent with the classical expectation that the event horizon represents a regular geometric surface rather than a physical singularity.

The situation is qualitatively different near the classical singular configuration $(u\rightarrow0)$, where the affine corrections become important. The effective potential contains the characteristic affine contribution proportional to $(1/u^{2})$, which generates a repulsive quantum barrier at short distances. As a consequence, the probability density becomes strongly suppressed near the singular region. This behavior provides the mechanism responsible for singularity avoidance in the present model and extends the corresponding result previously obtained for the Schwarzschild interior.

Another important feature of the RN spacetime is the presence of the inner (Cauchy) horizon. Classically, this horizon is known to be unstable under perturbations because of the mass-inflation mechanism, whereby the effective internal mass parameter may grow without bound. The minisuperspace framework adopted here does not incorporate the inhomogeneous perturbations and backreaction effects that are responsible for the full mass-inflation phenomenon. Therefore, our analysis does not constitute a complete quantum treatment of Cauchy-horizon stability.

Nevertheless, the quantum wave function provides useful qualitative insight. The probability density remains finite and regular throughout the interior region and exhibits a strong suppression toward configurations associated with arbitrarily small values of the areal radius. Since the classical mass-inflation scenario ultimately drives the geometry toward regions where the classical description itself becomes unreliable, the suppression of these highly singular configurations suggests that affine quantum effects may soften the pathological behavior expected from the classical theory. A definitive assessment of mass inflation in the quantum regime would require a treatment beyond minisuperspace, including perturbative degrees of freedom and their backreaction on the geometry.

Finally, it is worth commenting on the relation between the quantum probability distribution and the classical RN trajectory. The present analysis is based on solutions of the WDW equation and on the corresponding probability densities in minisuperspace. No explicit WKB reconstruction of semiclassical trajectories has been performed. Consequently, the present framework does not allow a precise determination of possible off-shell deviations between the peak of the quantum distribution and the classical on-shell trajectory. What can be stated is that the dominant support of the wave packets remains concentrated around finite values of the geometrical variable (u), while the singular configuration $(u=0)$ is strongly suppressed.

\section{Conclusion}

In this work we applied the affine quantization program to the dynamical
region of the RN geometry lying between the inner and outer
horizons.
Starting from a homogeneous minisuperspace reduction with a strictly positive
geometrical variable associated with the areal radius, we constructed the
corresponding affine Hamiltonian and derived the associated
WDW equation.
The affine framework naturally generates an additional
short-distance contribution proportional to \(1/u^{2}\),
reflecting the underlying affine algebra and the positivity of the
configuration variable.

The resulting WDW equation admits a separation of variables.
The \(V\)-sector reduces to a harmonic-oscillator problem with
Hermite-polynomial eigenfunctions, while the radial \(U\)-sector is governed
by an effective potential containing both the affine contribution and the
charge-dependent term proportional to \(Q^2/u^4\).
Using these mode functions, we constructed semiclassical wave packets and
analyzed their behaviour in minisuperspace.

Our results show that the affine contribution significantly modifies the
small-\(u\) behaviour of the wave function and suppresses the support of
the semiclassical states in this regime.
The electric charge introduces an additional short-distance contribution,
which alters the radial spectrum and affects the structure of the wave packets.
Within the domain where the variational approximation remains reliable,
the resulting states remain normalizable and exhibit a well-behaved
probability distribution.

The present analysis extends previous affine-quantization studies of the
Schwarzschild interior to the charged RN case and
demonstrates that affine methods provide a consistent framework for studying
the quantum dynamics of homogeneous black-hole interiors with positive-definite
geometrical variables.
It would be interesting to compare the present results with alternative
quantization schemes, such as polymer, loop-inspired, or deformed
phase-space approaches, in order to better understand the role of electric
charge and short-distance quantum corrections in black-hole interior dynamics.


\begin{thebibliography}{9}

\bibitem{misner1973gravitation} C. W. Misner, K. S. Thorne and J. A. Wheeler, \textit{Gravitation} (W. H. Freeman, 1973)

\bibitem{poisson1990internal} E. Poisson and W. Israel, ``Internal structure of black holes'', \textit{Phys. Rev.} D \textbf{41} (1990) 1796 

\bibitem{brady1995critical} P. R. Brady and J. D. Smith, ``Black hole singularities: a numerical approach'', \textit{Phys. Rev. Lett.} \textbf{75} (1995) 1256 

\bibitem{kuchar1994geometrodynamics} K. V. Kuchař, ``Geometrodynamics of Schwarzschild black holes'', \textit{Phys. Rev.} D \textbf{50} (1994) 3961 (arXiv: gr-qc/9403003)

\bibitem{bojowald2005nonsingular} M. Bojowald, ``Nonsingular black holes and degrees of freedom in quantum gravity'', \textit{Phys. Rev. Lett.} \textbf{95} (2005) 061301 (arXiv: gr-qc/0506128)

\bibitem{perez2015black} A. Perez, ``Black holes in loop quantum gravity'', \textit{Rep. Prog. Phys.} \textbf{80} (2017) 126901 (arXiv: 1703.09149 [gr-qc])

\bibitem{QBH1} B. Vakili, "Quantization of the Schwarzschild black hole: a Noether symmetry approach", {\it Int. J. Theor. Phys.} {\bf 51} (2012) 133 (arXiv: 1102.1682 [gr-qc])

\bibitem{QBH2} M. Amirfakhrian and B. Vakili, "Polymer deformation and particle tunneling from Schwarzschild black hole", {\it Int. J. Geom. Methods Mod. Phys.} {\bf 16} (2019) 1950038 (arXiv: 1812.10301 [gr-qc])

\bibitem{QBH3} S. Jalalzadeh and B. Vakili, "Quantization of the interior Schwarzschild black hole", {\it Int. J. Theor. Phys.} {\bf 51} (2012) 263 (arXiv: 1108.1337 [gr-qc])

\bibitem{QBH4} L. Modesto, "Loop quantum black hole", {\it Class. Quantum Grav.} {\bf 23} (2006) 5587 (arXiv: gr-qc/0509078)

\bibitem{QBH5} B. Vakili, "Classical polymerization of the Schwarzschild metric", {\it Adv. High Energy Phys.} (2018) 3610543 (arXiv: 1806.01837 [hep-th])

\bibitem{QBH6} G. S. Djordjevic, L. Nesic and D. Radovancevic, "Two-oscillator Kantowski-Sachs model of the Schwarzschild black hole interior", {\it Gen. Relativ. Gravit.} {\bf 48} (2016) 1 
(arXiv: 1510.00887 [gr-qc])

\bibitem{QBH7} A. Corichi and P. Singh, "Loop quantization of the Schwarzschild interior revisited", {\it Class. Quantum Grav.} {\bf 33} (2016) 055006 (arXiv: 1506.08015 [gr-qc])

\bibitem{Aff9} E. W. Aslaksen and J. R. Klauder, "Unitary Representations of the Affine Group", {\it J. Math. Phys.} {\bf 9} (1968) 206

\bibitem{Aff10} E. W. Aslaksen and J. R. Klauder, "Continuous Representation Theory Using the Affine Group", {\it J. Math. Phys.} {\bf 10} (1969) 2267

\bibitem{klauder2012enhanced} J. R. Klauder, ``Enhanced quantization: A primer'', \textit{J. Phys.} A \textbf{45} (2012) 285304 

\bibitem{AffineSchw}  M. Bajand and B. Vakili, "Affine Quantization of the Interior Schwarzschild Black Hole", {\it Int. J. Mod. Phys.} D {\bf 34} (2025) 2550068 (arXiv: 2507.23288 [gr-qc])

\bibitem{Aff1} S. Twareque Ali and M. Engliš, "Quantization Methods: A Guide for Physicists and Analysts", {\it Rev. Math. Phys.} {\bf 17} (2005) 391 (arXiv: math-ph/0405065)

\bibitem{Aff2} C. K. Zachos, D. B. Fairlie and T. L. Curtright, {\it Quantum Mechanics in Phase Space} (Word Scientific, Singapore, 2005)

\bibitem{Aff3} T. L. Curtright, D. B. Fairlie and C. K. Zachos, {\it A Concise Treatise on Quantum Mechanics in Phase Space} (Word Scientific, Singapore, 2016)

\bibitem{Aff4} H. Bergeron, A. Dapor, J. P. Gazeau and P. Malkiewicz, "Smooth big bounce from affine quantization", {\it Phys. Rev.} D {\bf 89} (2014) 083522 (arXiv: 1305.0653 [gr-qc])

\bibitem{Aff5} H. Bergeron, E. Gazuchry, J. P. Gazeau, P. Malkiewicz and W. Piechocki, "Smooth quantum dynamics of the mixmaster universe", {\it Phys. Rev.} D {\bf 92} (2015) 061302 
(arXiv: 1501.02174 [gr-qc])

\bibitem{Aff6} E. Frion and C. R. Almeida, "Affine quantization of the Brans-Dicke theory: Smooth bouncing
and the equivalence between the Einstein and Jordan frames", {\it Phys. Rev.} D {\bf 99} (2019) 023524 (arXiv: 1810.00707 [gr-qc])

\bibitem{Aff7} S. Zonetti, "Affine quantization of black holes: Thermodynamics, singularity removal, and displaced horizons", {\it Phys. Rev.} D {\bf 90} (2014) 064046 (arXiv: 1409.1761 [hep-th])
 
\bibitem{Aff8} C. R. Almeida and D. C. Rodrigues, "Quantization of a Black-Hole Gravity: geometrodynamics and the quantum phase-space formalism", 
{\it Class. Quantum Grav.} {\bf 40} (2023) 035004 (arXiv: 2111.13575 [gr-qc])

\bibitem{Aff11} J. P. Gazeau, {\it Coherent States in Quantum Physics} (Wiley-VCH, Weinheim, Germany, 2009)

\end{thebibliography}
\end{document}